\documentclass[lettersize,journal]{IEEEtran}
\pdfoutput=1
\usepackage{amsmath,amsfonts}
\usepackage{algorithmic}
\usepackage{algorithm}
\usepackage{array}
\usepackage{textcomp}
\usepackage{stfloats}
\usepackage{verbatim}
\usepackage{graphicx}
\usepackage{cite}
\usepackage{subfigure}
\usepackage{booktabs}
\usepackage{framed,multirow}
\usepackage{amssymb}
\usepackage{latexsym}
\usepackage{xcolor}
\usepackage{bbding}
\usepackage{caption}
\usepackage{float}

\begin{document}

\title{The state-of-the-art 3D anisotropic intracranial hemorrhage segmentation on non-contrast head CT: The INSTANCE challenge}

\author{Xiangyu Li, Gongning Luo, Kuanquan Wang, Hongyu Wang, Jun Liu, Xinjie Liang, Jie Jiang, Zhenghao Song, Chunyue Zheng, Haokai Chi, Mingwang Xu, Yingte He, Xinghua Ma, Jingwen Guo, Yifan Liu, Chuanpu Li, Zeli Chen, Md Mahfuzur Rahman Siddiquee, Andriy Myronenko, Antoine P. Sanner, Anirban Mukhopadhyay, Ahmed E. Othman, Xingyu Zhao, Weiping Liu, Jinhuang Zhang, Xiangyuan Ma, Qinghui Liu, Bradley J MacIntosh, Wei Liang, Moona Mazher, Abdul Qayyum, Valeriia Abramova, Xavier Lladó, Shuo Li}
   

\markboth{Journal of \LaTeX\ Class Files,~Vol.~14, No.~8, August~2021}%
{Shell \MakeLowercase{\textit{et al.}}: A Sample Article Using IEEEtran.cls for IEEE Journals}


\maketitle

\begin{abstract}
Automatic intracranial hemorrhage segmentation in 3D non-contrast head CT (NCCT) scans is significant in clinical practice. Existing hemorrhage segmentation methods usually ignores the anisotropic nature of the NCCT, and are evaluated on different in-house datasets with distinct metrics, making it highly challenging to improve segmentation performance and perform objective comparisons among different methods. The 2022 intracranial hemorrhage segmentation on non-contrast head CT (INSTANCE 2022) was a grand challenge held in conjunction with the 2022 International Conference on Medical Image Computing and Computer Assisted Intervention (MICCAI). It is intended to resolve the above-mentioned problems and promote the development of both intracranial hemorrhage segmentation and anisotropic data processing. The INSTANCE released a training set of 100 cases with ground-truth and a validation set with 30 cases without ground-truth labels that were available to the participants. A held-out testing set with 70 cases is utilized for the final evaluation and ranking. The methods from different participants are ranked based on four metrics, including 
Dice Similarity Coefficient (DSC), Hausdorff Distance (HD), Relative Volume Difference (RVD) and Normalized Surface Dice (NSD). A total of 13 teams submitted distinct solutions to resolve the challenges, making several baseline models, pre-processing strategies and anisotropic data processing techniques available to future researchers. The winner method achieved an average DSC of 0.6925, demonstrating a significant growth over our proposed baseline method. To the best of our knowledge, the proposed INSTANCE challenge releases the first intracranial hemorrhage segmentation benchmark, and is also the first challenge that intended to resolve the anisotropic problem in 3D medical image segmentation, which provides new alternatives in these research fields. 
\end{abstract}


\section{Introduction}
\IEEEPARstart{I}{ntracranial} hemorrhage (ICH) is a severe brain disease and a main cause of stroke \cite{caceres2012intracranial,morotti2020noncontrast}. It has 
a high mortality rate of $40\%$ within one month \cite{van2010incidence,heit2017imaging}. Furthermore, ICH even causes significant disability in survivor patients, with only 20\% of patients expected to be capable of living independently in half year \cite{goldstein2011critical}. Therefore, early and accurate diagnosis of the ICH is important for saving patients' lives and improve their prognosis in clinical practice\cite{caceres2012intracranial, li2021hematoma}. Non-contract head computerized tomography (NCCT) is the primary imaging modality to diagnosing ICH for its widely availability in most emergency rooms and high sensitivity for detecting ICH. Moreover, NCCT enables accurate monitoring of hemorrhage progression, and effectively quantify hematoma volumes in ICH \cite{macellari2014neuroimaging,caceres2012intracranial,heit2017imaging}, making it a gold standard examination for the diagnosis of ICH. 

Hematoma volume estimation is significant for the prognosis and treatment decisions for ICH patients. In recent clinical trials, the hematoma volume has been utilized as a reliable indicator to determine the optimal candidates for intervention \cite{ironside2020fully,hanley2019efficacy,broderick1993volume}. Thus, 
volume quantification of ICH has become an essential procedure for outcome predictions and ICH therapy. The hematoma volume can be estimated by semiautomated methods with the aid of 
radiologists, which is time-consuming \cite{prakash2012segmentation} and suffers from inter-rater variability \cite{islam2018ichnet}. The ABC/2 method \cite{kothari1996abcs} is an effective technique to estimate hematoma volume in clinical practice since it is simple to implement. However, the estimation accuracy of the ABC/2 method dramatically decreases with irregular or large scale hemorrhages \cite{webb2015accuracy, ironside2020fully}. 
The ICH segmentation methods, enabling accurate and rapid hematoma volume quantification, have become the leading criterion in ICH diagnosis. 

However, there exists plenty of challenges to segment ICH for automatic methods. For example, the hemorrhage structures vary considerably across patients in terms of shape, size, and localization, preventing the use of valuable location and shape priors that are significant elements in the segmentation of many other anatomical structures. The blurred boundaries for the ICH region further improve the difficulty of the segmentation task \cite{cho2019improving}.

Because of the clinical significance and the intrinsic challenges, the task of automatic intracranial hemorrhage segmentation has attracted extensive attention in the past few years. Recently, deep learning–based ICH segmentation models that segment ICH regions and quantify hematoma volume have been performed to effectively diagnose ICH and have achieved competitive results \cite{kyung2022improved, toikkanen2021resgan, li2021hematoma, chang2018hybrid, patel2019intracerebral, kwon2019siamese}. However, all those above-mentioned ICH segmentation methods ignore the anisotropic nature of the NCCT volume by simply performing 2D or 3D convolutional networks, and they were evaluated on different in-house hemorrhage segmentation datasets with distinct metrics, making it highly challenging to improve segmentation performance and perform objective comparisons among these methods. Consequently, it remains hard to determine which kinds of segmentation techniques may be valuable to follow in clinical practice and research; what exactly the performance is of the state-of-the-art methods.

To resolve the above-mentioned challenges on fair comparisons of different methods, we organized the \textbf{IN}tracranial hemorrhage \textbf{S}egmen\textbf{TA}tio\textbf{N} \textbf{C}halleng\textbf{E} on non-contrast head CT (\textbf{INSTANCE}) in conjunction with the 2022 international conference on Medical Image Computing and Computer Assisted Interventions (MICCAI) in Singapore. To this end, we collected and released an ICH segmentation dataset of 200 3D volumes with refined labeling from several experienced radiologists, and encouraged the participants to develop novel algorithms to effectively segment hematoma region with anisotropic NCCT volumes. Moreover, we evaluate different benchmark ICH segmentation methods with the same metrics, including Dice Similarity Coefficient (DSC), Hausdorff distance (HD), relative volume difference (RVD) and normalized surface dice (NSD). Each of these benchmark methods was implemented by different challenge participants on a subset of the ICH dataset, and tested on a isolated testing dataset against the manually delineated groundtruth labels. To the best of our knowledge, INSTANCE is the first public intracranial hemorrhage segmentation challenge, and also the first challenge that intended to deal with the anisotropic problem in 3D biomedical image segmentation. It is served as a solid benchmark for ICH segmentation tasks, and would also promote the development of intracranial hemorrhage segmentation and anisotropic data processing.

\section{Prior Works}

\subsection{Related intracranial hemorrhage segmentation methods}
A large numbers of methods have been proposed to automatically segment intracranial hemorrhage in CT scans. Among them, deep learning techniques are widely adopted for its excellent performance in medical image segmentation tasks \cite{lee2019explainable,cho2019improving}. Ironside et al. utilized U-Net  \cite{ronneberger2015u} to automatically segment ICH and estimate the hematoma volume. They achieved comparable accuracy and greater efficiency compared to manual and semi-automated segmentation techniques \cite{ironside2020fully}. To address the issue of insufficient annotation data for ICH segmentation tasks, Kuo et al. proposed a patch-based FCN network and segmented ICH in an active learning manner \cite{kuo2018patchfcn}. Chang et al. proposed an ROI-based framework that is optimized specifically for ICH detection and segmentation tasks by projecting 3D features to 2D networks in the feature pyramid network  \cite{chang2018hybrid}. Kwon et al. proposed a Siamese U-Net method to segment ICH by leveraging the dissimilarity between learned features of healthy templates and input images \cite{kwon2019siamese}. Kyung et al. proposed a supervised multi-task aiding representation transfer learning network for ICH, which was divided into upstream and downstream. In the upstream, effective representation learning was performed by multi-task learning (classification, segmentation, reconstruction) and differences in the specific head of the consistency loss mitigation target are added. For downstream, feature extractor trained upstream is combined with 3D operator (classifier or divider) to implement specific tasks \cite{kyung2022improved}. Wu et al. proposed a combination of an attention-based convolutional neural network and a variational Gaussian process for multiple instance learning method for predicting intracranial hemorrhage slices \cite{wu2021combining}. Toikkanen et al. proposed a residual segmentation method based on generative adversarial network, which generates the image without bleeding in the original section through the model, and then calculates the difference between the generated image and the original image, so as to obtain the segmented image \cite{toikkanen2021resgan}. Abramova et al. introduced the squeeze-excitation block into 3D U-Net to solve the problem of segment hemorrhagic stroke lesions. Moreover, a restrictive patch sampling is proposed to alleviate the class imbalance problem and also to deal with the issue of intra-ventricular hemorrhage \cite{abramova2021hemorrhagic}. Kuang et al. designed new self-attention blocks and contextual attention blocks that take full advantage of both in-chip and inter-chip information. In addition, multilevel training strategies are proposed to reduce the influence of inter-class imbalance \cite{kuang2020psi}. Wang et al. propose a Masked Multi-Task Network method to detect brain CT volumes with intracranial hemorrhage and distinguish hemorrhage type by leveraging different types of intracranial hemorrhage at different locations \cite{wang2020masked}. Guo et al. propose a full convolutional neural network for simultaneous classification and segmentation of ICH, and the ConvLSTM module was used to address this issue of the loss of spatial information \cite{guo2020simultaneous}. Kadam et al. propose architectures combined Xception and LSTM/GRU for classification of Intracranial Hemorrhage. It is also found through experiments that Xception GRU model has better performance  on most of the metrics as compared to the Xception and Xception LSTM models \cite{kadam2021cnn}.

Despite the excelent results reported in the above papers, it is still challenging to identify the best performing method among them because of the varied testing datasets and evaluation metrics.
The proposed INSTANCE challenge provides a standardized procedure to systematically evaluate and compare different SOTA methods on the same testing dataset and consistent evaluation metrics, enabling objective and fair comparison among different techniques.

\subsection{Medical Image Segmentation Challenges}
Recently years have witnessed the growing popularity for biomedical image analysis challenges, especially for medical image segmentation challenges. To name a few, there were 25, 20, and 40 accepted challenges at the International Conference on Medical Image Computing and Computer-Assisted Intervention (MICCAI) 2020, 2021, and 2022, respectively. From 2020 to 2022, the number of challenges nearly doubled, and the segmentation-related challenges occupied $38\%$ of all the challenges\footnote{https://www.biomedical-challenges.org/}. Similarly, in the largest biomedical image challenge platform 'Grand Challenge\footnote{https://grand-challenge.org/}', 149 out of 315 ($47.3\%$) challenges are designed for segmentation tasks. There are lots of successful challenges in medical image segmentation, for example, the Brain Tumor Segmentation (BraTS) challenge \cite{menze2014multimodal} provide a solid benchmark for multimodal brain tumor segmentation task, numerous methods on brain tumor segmentation and multi-modal learning have been validated on this benchmark, significantly improving the development of those research fields. The Head and neck tumor segmentation challenge (Hecktor) \cite{oreiller2022head} organized a novel challenge for head and neck tumor segmentation on PET/CT modalities, which claimed to be the pioneer work on this field.   
The abdomen ct organ segmentation \cite{ma2022fast} first consider the inference time, and GPU memory consumption as extra evaluation metrics instead of simply focusing on the segmentation accuracy, providing a novel benchmark with more comprehensive evaluation metrics. The Kidney Tumor Segmentation (KiTS) ( \cite{heller2021state} ) challenge allow participants to compare their methods on kidney and kidney tumor segmentation tasks.

Those above challenges have made great progress in promoting the development of specific medical field. However, to the best of our knowledge, there are no challenges intended to resolve the ICH segmentation with anisotropic 3D volumes. Hence, the INSTANCE is the first released grand challenge for the ICH segmentation and also the first challenge that intended to deal with the anisotropic problem in 3D medical image segmentation. We believe that the ICH data and algorithms provided in this benchmark would be helpful to promote the development of both ICH diagnosis and anisotropic data processing. 

\section{The Organization of the INSTANCE challenge}
The proposed INSTANCE challenge was organized in 2022 and was in conjunction with the 25rd MICCAI conference as a satellite event. It was deployed on the Grand Challenge platform. The official webpage of the INSTANCE challenge is \url{ https://instance.grand-challenge.org/}. Meanwhile, we also construct a Github repository \footnote{https://github.com/PerceptionComputingLab/INSTANCE2022} which provides plenty of resources related to the challenge, for example, the agreement files for accessing the dataset, the docker rules and submission examples, and also the baseline models. For the challenge schedule, the registration is open to the public on March 28, 2022. The training and validation dataset were released on April 6 and July 15, respectively. The deadline of the open validation phase and the testing phase is on August 7 and August 14, respectively. In the validation phase, the participants uploaded their segmentation results to the Grand challenge website, and the platform automatically calculated the evaluation metrics by comparing them with the ground-truth labels we provided, and then displayed the calculated metrics on the validation leaderboard \footnote{https://instance.grand-challenge.org/evaluation/challenge/leaderboard/}
In the testing phase, the participants are required to submit one successful docker image that contains their algorithms, and we ran the docker images from different participants on the closed testing dataset. The dataset of the INSTANCE challenge are currently available to the public on Grand Challenge platform after signing an agreement file and the post-challenge leaderboard submission is open for researches in this community. The following sections summarizes the detailed implementation of the INSTANCE challenge. 
\subsection{Dataset}
\begin{figure*}
	\centering
	\includegraphics[height=4.048cm,width=18cm]{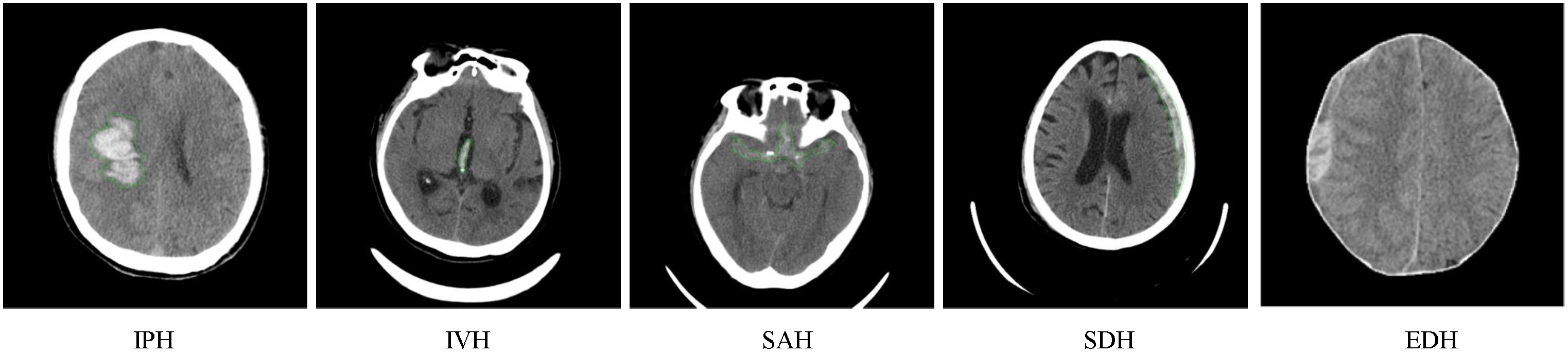}
	\caption[]{Different kinds of intracranial hemorrhages, including intraparenchymal hemorrhage (IPH), intraventricular hemorrhage (IPH), subarachnoid hemorrhage (SAH), subdural hemorrhage (SDH), and epidural hemorrhage (EDH). The varied shapes and positions for different kinds of hemorrhages promote the difficulties of the segmentation task.} 
	\label{fig:ich_class}
\end{figure*}

We obtained the approval from Peking university, shougang hospital to perform a retrospective analysis of the patients that were diagnosed as intracranial hemorrhage between 2017 and 2019. We then collected 200 non-contrast head CT volumes of those patients to construct challenge dataset. For these 200 cases, they were diagnosed as  different kinds of ICHs, including intraparenchymal hemorrhage (IPH), intraventricular hemorrhage (IPH), subarachnoid hemorrhage (SAH), subdural hemorrhage (SDH), and epidural hemorrhage (EDH), an example for each kind of ICH is illustrated in Fig. \ref{fig:ich_class}. We then split the 200 cases into training, validation and testing, with 100, 30, and 70 cases respectively. The CT scans and the labels of the training set are available to the participant for model training, while only the CT scans are provided for them to tune their algorithms on the Grand Challenge platform. Finally, in the testing phase, we provide a closed test set for fair comparison between different methods.

For each of the subject in INSTANCE dataset, we first converted the traditional 
Digital Imaging and Communications in Medicine (DICOM) files to the Neuroimaging Informatics Technology Initiative (NIfTI) format. In this way, each subject only has one single NIfTI file instead of a bunch of DICOM files, making it easier to process in a image segmentation program. The volume sizes ranges from $512\times512\times20$ to $512\times512\times70$, and the pixel spacing of a CT volume is $0.42mm\times0.42mm\times5mm$, hence the volume is anisotropic with inter-slice resolution much smaller than the within-slice resolution. The window width and the window center is $90HU$ and $40HU$, respectively. We kept the original Hu value in the NIfTI volume since the participants can conduct different windowing strategies. 

For the data annotation, we gathered several experienced radiologists and some postgraduate students majored in medical imaging to perform hemorrhage region annotation in the NCCT scans. To improve the efficiency of the annotation process, we adopted a coarse-to-fine annotation strategy. Specifically, the ICH lesions were first manually delineated in the NCCT volumes with a popular annotation software in medical imaging, Seg3D\footnote{https://www.sci.utah.edu/cibc-software/seg3d.html} \cite{SCI:Seg3D}. Then the experienced radiologists checked the coarse annotations and manually refined them. Finally, all the radiologists double-check the annotations from other annotators, and discuss to achieve the final annotations with majority voting strategy. 
\subsection{Evaluation Measures and Ranking Method}\label{subsec:eval_metrics}
The INSTANCE challenge adopted four accuracy-related evaluation metrics: Dice Similarity Coefficient (DSC), Hausdorff Distance (HD), Relative Volume Difference (RVD) and Surface Dice (NSD) \cite{nikolov2021clinically}. We utilized DSC and HD since they are widely used in different medical image segmentation challenges. They are complementary metrics for evaluating segmentation performance. DSC was utilized to measure the region overlapping error between ground truth and segmentation results, while HD is used to evaluate the coincidence between segmented surface and target surface. We used the RVD since the purpose for the ICH segmentation is to quantify the hematoma volume, making the volume differences between the predictions and the labels significant for the results analysis. Moreover, we further added the NSD metric as a complement evaluation for the HD metric because the HD would become infinite when the prediction is a normal head CT scan without hemorrhages. The NSD also measures the discrepancy between the target and predicted boundaries.  

We intended to rank different algorithms based on the above-mentioned four metrics. Motivated from the former challenges \cite{lalande2022deep, oreiller2022head}, we utilized a “aggregate-then-rank” scheme for ranking, including the following three steps: (1) Calculate the average DSC, HD, RVD and NSD metrics for all cases in the testing dataset. (2) Rank all the participant teams on these four metrics, hence each team would get four ranks. (3)Based on the rankings generated from (2), we then averaged these rankings and achieved the final ranking for each team. Moreover, for some extreme cases, e.g., the HD metric is infinite because the algorithm mistakenly treated some hard ICH cased as normal head scans. In this case, we treat all  ‘inf’ teams the same rank on HD which are inferior to others. Because we believe effectively diagnosis hard samples is also important in our challenge.
\section{Results}
\begin{table}
	\centering
	\caption{The Correspondence between the Team names and the aliases.}
	\label{tab:team_name} 
	\tabcolsep=5px
	{
		\begin{tabular}{cc}
			\hline\noalign{\smallskip}	
			Team & Alias  \\
			\noalign{\smallskip}\hline\noalign{\smallskip}
			vegetable&T1\\
			nvauto&T2\\
			mec-lab&T3\\
			ibot&T4\\
			stubmers&T5\\
			crainet&T6\\
			superembrace&T7\\
			scan&T8\\
			dolphins&T9\\
			nic-vicorob&T10\\
			2i\_mtl&T11\\
			avich&T12\\
			visal&T13\\
			\noalign{\smallskip}\hline
	\end{tabular}}
\end{table}
\begin{table*}
	\centering
	\caption{Summary of the algorithms in terms of key factors in the methods by those participants: backbone network, 2D/3D, stages, pre-processing, data augmentation, loss functions, ensembles, post-processing. Abbreviation: Normalization (N), Windowing(W), Skull stripping(SS), Cropping (CP), Cross-Entropy(CE), Tversky (TV), Contour loss (CT) }
	\label{tab:algorithm_summary} 
	\tabcolsep=5px
	{
		\begin{tabular}{llllllllll}
			\hline\noalign{\smallskip}	
			Team & Backbone & 2D/3D & Preprocess & Stage & Augmentation & Loss & Ensemble & Postprocess & Patch-based \\
			\noalign{\smallskip}\hline\noalign{\smallskip}
			T1 & nnU-Net & 2D/3D & N & 1 & $\checkmark$ & Dice+WCE & $\checkmark$ & \XSolidBrush &  $\checkmark$ \\
			T2 & ResUNet & 2D & N & 2 & $\checkmark$ & Dice+CE & $\checkmark$ & \XSolidBrush &  \XSolidBrush\\
			T3 & nnU-Net & 3D & SS & 1 & $\checkmark$ & Dice+CE+CT & $\checkmark$ & \XSolidBrush &  \XSolidBrush\\
			T4 & nnU-Net & 3D & N+W+CP & 1 & $\checkmark$ & Dice+CE & \XSolidBrush & \XSolidBrush &  $\checkmark$\\
			T5 & nnU-Net & 3D & N+SS & 1 & $\checkmark$ & Dice+CE & $\checkmark$ & \XSolidBrush &  $\checkmark$\\
			T6 & nnU-Net & 3D & N+W & 2 & $\checkmark$ & Dice+Focal & $\checkmark$ & \XSolidBrush &  $\checkmark$\\
			T7 & nnU-Net & 3D & N+W & 1 & $\checkmark$ & Dice+CE & $\checkmark$ & \XSolidBrush &  $\checkmark$\\
			T8 & U-Net & 3D & W & 1 & $\checkmark$ & CE & $\checkmark$ & \XSolidBrush &  $\checkmark$\\
			T9 & nnU-Net & 2D/3D & N & 2 & $\checkmark$ & CE & $\checkmark$ & \XSolidBrush &  $\checkmark$\\
			T10 & U-Net & 3D & N+SS & 1 & $\checkmark$ & Dice+CE & $\checkmark$ & $\checkmark$ &  $\checkmark$\\
			T11 & Attention U-Net & 2D & N & 2 & $\checkmark$ & Dice+CE+TV & $\checkmark$ & \XSolidBrush &  \XSolidBrush\\
			T12 & U-Net & 2D & N & 1 & $\checkmark$ & Dice & \XSolidBrush & \XSolidBrush &  \XSolidBrush\\	
			T13 & U-Net3+ & 2D & SS & 1 & $\checkmark$ & Dice+CE & \XSolidBrush & \XSolidBrush &  \XSolidBrush\\
			\noalign{\smallskip}\hline
	\end{tabular}}
\end{table*}
\begin{table*}
	\centering
	\caption{Summary of the INSTANCE 2022 validation phase. The average DSC, RVD, NSD and HD are reported for the baseline models and the submitted algorithms from each participant. The unit of HD is [mm]. Bold values represent the best scores for each metric.}
	\label{tab:challenge_results_validation} 
	\tabcolsep=5px
	{
		\begin{tabular}{cccccc}
			\hline\noalign{\smallskip}	
			Team & DSC(\%)$\uparrow$ & NSD(\%)$\uparrow$ & RVD$\downarrow$ & HD$\downarrow$\\
			\noalign{\smallskip}\hline\noalign{\smallskip}
			T1 & 79.12$\pm$23.00 & 50.26$\pm$19.91 & 0.21$\pm$0.20 & 29.02$\pm$26.34\\
			T2 & 78.21$\pm$18.45 & 55.28$\pm$12.67 & \textbf{0.20$\pm$0.18} & 32.30$\pm$30.04\\
			T3 & 71.60$\pm$30.10 & 50.60$\pm$21.30 & 0.29$\pm$0.30 & inf\\
			T4 & 73.55$\pm$26.74 & 51.57$\pm$18.10 & 0.24$\pm$0.24 & 27.16$\pm$32.41\\
			T5 & 73.39$\pm$27.38 & 51.93$\pm$18.99 & 0.25$\pm$0.27 & inf\\
			T6 & \textbf{79.53 $\pm$17.18} & \textbf{56.81$\pm$12.47} & \textbf{0.20$\pm$0.18} & \textbf{21.56$\pm$25.02}\\
			T7 & 71.12$\pm$29.38 & 50.19$\pm$20.56 & 0.27$\pm$0.30 & inf\\
			T8 & 72.34$\pm$28.52 & 48.93$\pm$19.57 & 0.58$\pm$1.65 & 35.37$\pm$29.53\\
			T9 & 69.96$\pm$30.26 & 48.75$\pm$19.66 & 0.26$\pm$0.27 & inf\\
			T10 & 69.28$\pm$28.39 & 46.34$\pm$19.54 & 0.36$\pm$0.44 & 36.23$\pm$2.01\\
			T11 & 52.87$\pm$29.66 & 27.36$\pm$14.38 & 2.16$\pm$4.86 & 149.77$\pm$44.52\\
			T12 & 64.76$\pm$31.42 & 40.26$\pm$19.93 & 0.52$\pm$0.76 & 57.13$\pm$22.53\\
			T13 & 67.16$\pm$33.19 & 45.58$\pm$22.35 & 0.27$\pm$0.29 & 38.88$\pm$39.56\\
			\noalign{\smallskip}\hline
			Baseline \cite{li2021hematoma} & 64.08$\pm$27.48 & 46.21$\pm$20.12 & 0.514$\pm$1.14 &  277.63$\pm$163.00 \\
			\noalign{\smallskip}\hline
	\end{tabular}}
\end{table*}
\subsection{Participation and submissions}
The INSTANCE 2022 received over 500 applications on grand-challenge platform and 70 teams were approved to be able to access the challenge dataset. The reason why we refused the other applications was that they didn't submit the signed agreement files that we provided in the participation rules. In the validation phase, 30 teams uploaded their results with over 350 valid submissions on the grand challenge website. The final validation leaderboard is available on Grand challenge website. In the testing phase, 13 teams successfully submitted the Docker containers and the short papers.
\subsection{Algorithm summary}\label{subsec:algorithm_summary}
We adopted the SLEX-NET \cite{li2021hematoma} as the baseline model in the proposed INSTANCE challenge. It is noted that the dataset utilized in the SLEX-NET is different from INSTANCE 2022. Therefore, we re-trained the algorithm of baseline model on the INSTANCE 2022 dataset, with other training details consistent with the settings in the original paper. 

For the participants’ models, we find out that all the participants chose U-Net-related architectures, including attention U-Net \cite{oktay2018attention}, U-Net \cite{ronneberger2015u}, 3D U-Net\cite{cciccek20163d}, nnU-Net\cite{isensee2021nnu}, etc. Among them, nnUNet is still the most popular model, 7 out of 13 teams adopted it as their backbone network. Moreover, we also summarized other key factors in the methods by those participants, including data augmentation, loss functions, pre-processing, post-processing, and etc. The detailed summaries are illustrated in Table \ref{tab:algorithm_summary}. It shows that all teams used data augmentation, and 10 out of 13 teams conducted ensemble learning to improve their performance. In addition, four teams utilized the 2D implementation, seven teams adopted the 3D implementation, and two teams combined 2D/3D implementations. For the pre-processing and post-processing, all teams conducted different kinds of pre-processings, including normalization, windowing, skull-stripping, and etc, while only one team applied post-processing. To improve the learning of deep models, each team utilized different losses, such as Dice loss, cross-entropy loss, focal loss, and etc. Detailed descriptions of their methods can be found in the Appendix \ref{sec:appendix}. More importantly, we also released their submitted papers on the official challenge website \footnote{https://instance.grand-challenge.org/results/} for comprehensive introduction of their methods. 
\begin{table*}
	\centering
	\caption{Summary of the INSTANCE 2022 testing phase. The average DSC, RVD, NSD and HD are reported for the baseline models and the submitted algorithms from each participant. The unit of HD is [mm]. The ranking is only provided for teams that successfully submitted the docker image and the technical paper descriptions in the testing phase. Bold values represent the best scores for each metric.}
	\label{tab:challenge_results} 
	\tabcolsep=5px
	{
		\begin{tabular}{cccccc}
			\hline\noalign{\smallskip}	
			Team & DSC(\%)$\uparrow$ & NSD(\%)$\uparrow$ & RVD$\downarrow$ & HD$\downarrow$  &Ranking \\
			\noalign{\smallskip}\hline\noalign{\smallskip}
			T1 & 69.25$\pm$19.14 & \textbf{53.59$\pm$15.65} & \textbf{0.21$\pm$0.20} & \textbf{35.27$\pm$28.47} & 1\\
			T2 & \textbf{72.06$\pm$21.07} & 53.43$\pm$16.45 & 0.26$\pm$0.25 & inf & 2\\
			T3 & 69.00$\pm$24.68 & 51.25$\pm$18.94 & 0.31$\pm$0.28 & inf & 3\\
			T4 & 68.94$\pm$25.06 & 50.36$\pm$19.35 & 0.32$\pm$0.29 & inf & 4\\
			T5 & 67.97$\pm$25.07 & 49.46$\pm$18.73 & 0.32$\pm$0.28 & inf & 5 \\
			T6 & 67.39$\pm$26.91 & 48.40$\pm$20.84 & 0.32$\pm$0.31 & inf & 6\\
			T7 & 66.84$\pm$24.75 & 48.09$\pm$18.68 & 0.33$\pm$0.27 & 43.90$\pm$33.78 & 7\\
			T8 & 65.28$\pm$27.98 & 47.49$\pm$21.70 & 0.37$\pm$0.31 & inf & 8\\
			T9 & 64.97$\pm$26.78 & 46.86$\pm$19.58 & 0.34$\pm$0.31 & inf & 9\\
			T10 & 62.14$\pm$27.70 & 42.01$\pm$19.88 & 0.33$\pm$0.30 & inf & 10\\
			T11 & 61.95$\pm$25.91 & 42.80$\pm$18.75 & 0.40$\pm$0.78 & 55.36$\pm$26.53 & 11\\
			T12 & 57.04$\pm$28.20 & 36.73$\pm$19.17 & 0.43$\pm$0.50 & 60.81$\pm$25.22 & 12\\
			T13 & 40.22$\pm$32.35 & 25.11$\pm$21.54 & 1.55$\pm$4.67 & 68.36$\pm$41.79 & 13\\
			\noalign{\smallskip}\hline
			Baseline \cite{li2021hematoma} & 52.83$\pm$28.92 & 38.42$\pm$21.04 & 0.725$\pm$2.06 &  309.06$\pm$287.31 \\
			\noalign{\smallskip}\hline
	\end{tabular}}
\end{table*}
\begin{figure*}[htbp]
	\centering
	\subfigure[Dice Coefficient]{
		\includegraphics[width=8cm]{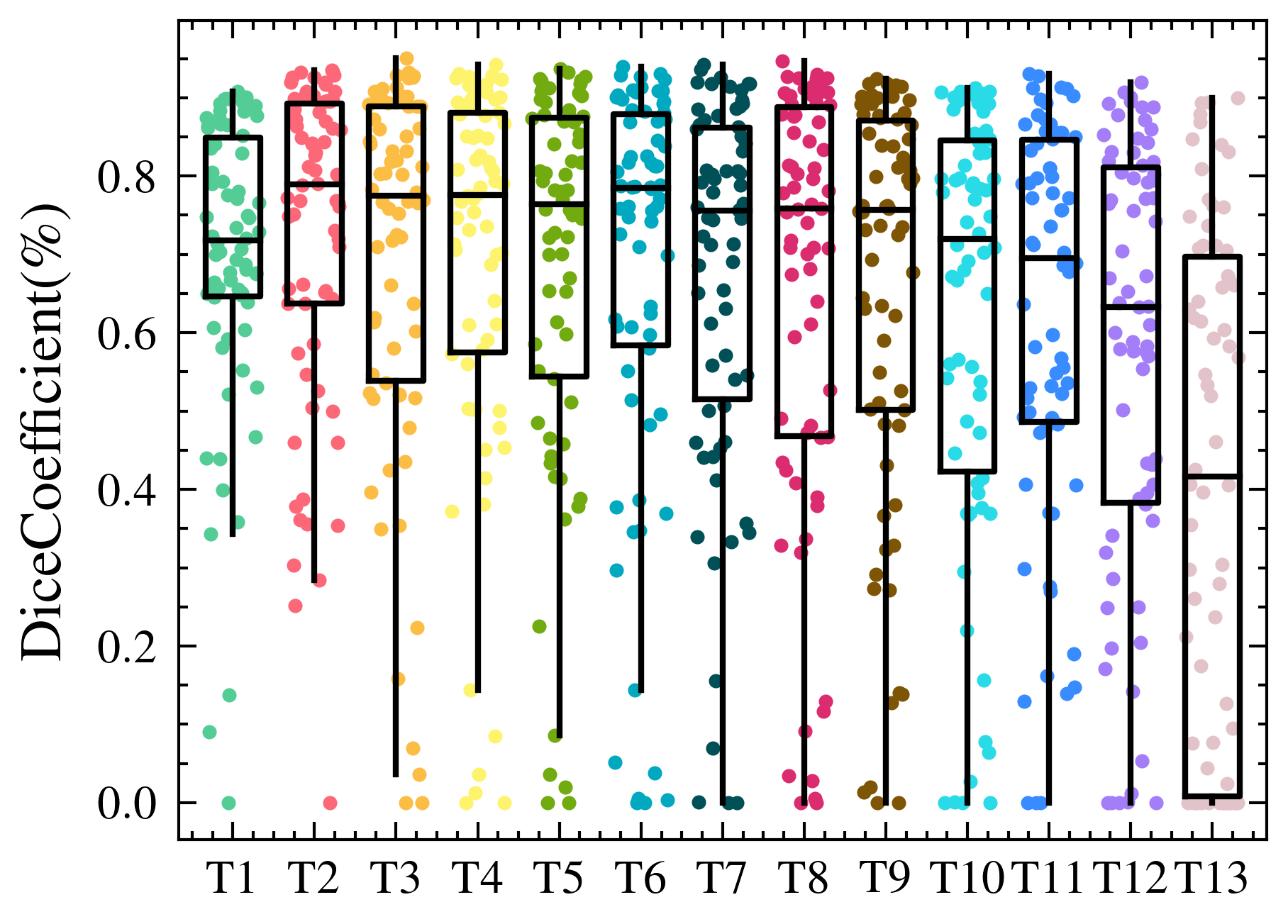}
		\label{subfig:dice}
	}
	\subfigure[Normalized Surface Dice]{
		\includegraphics[width=8cm]{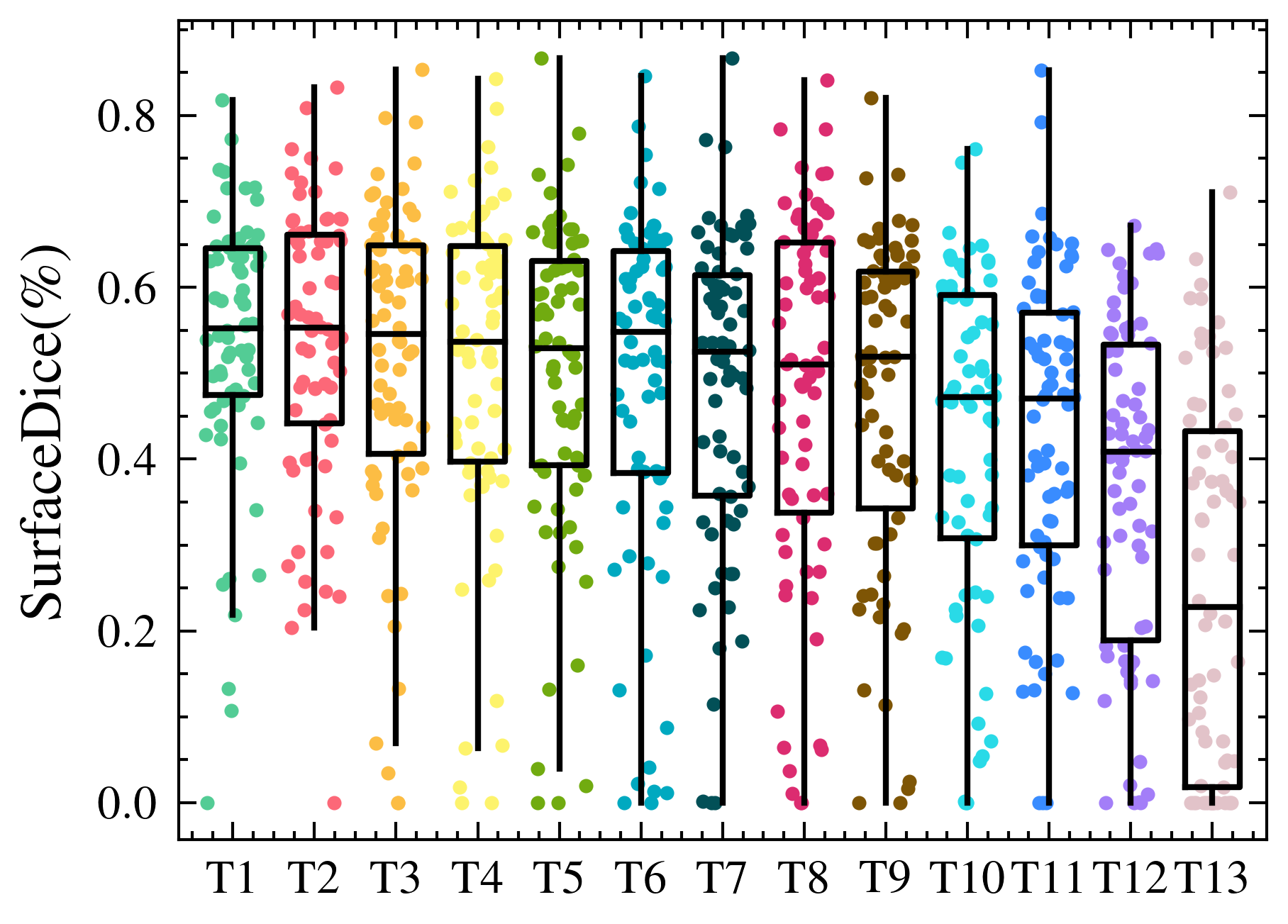}
		\label{subfig:SurfaceDice}
	}
	\quad  
	\subfigure[Relative Volume Difference]{
		\includegraphics[width=8cm]{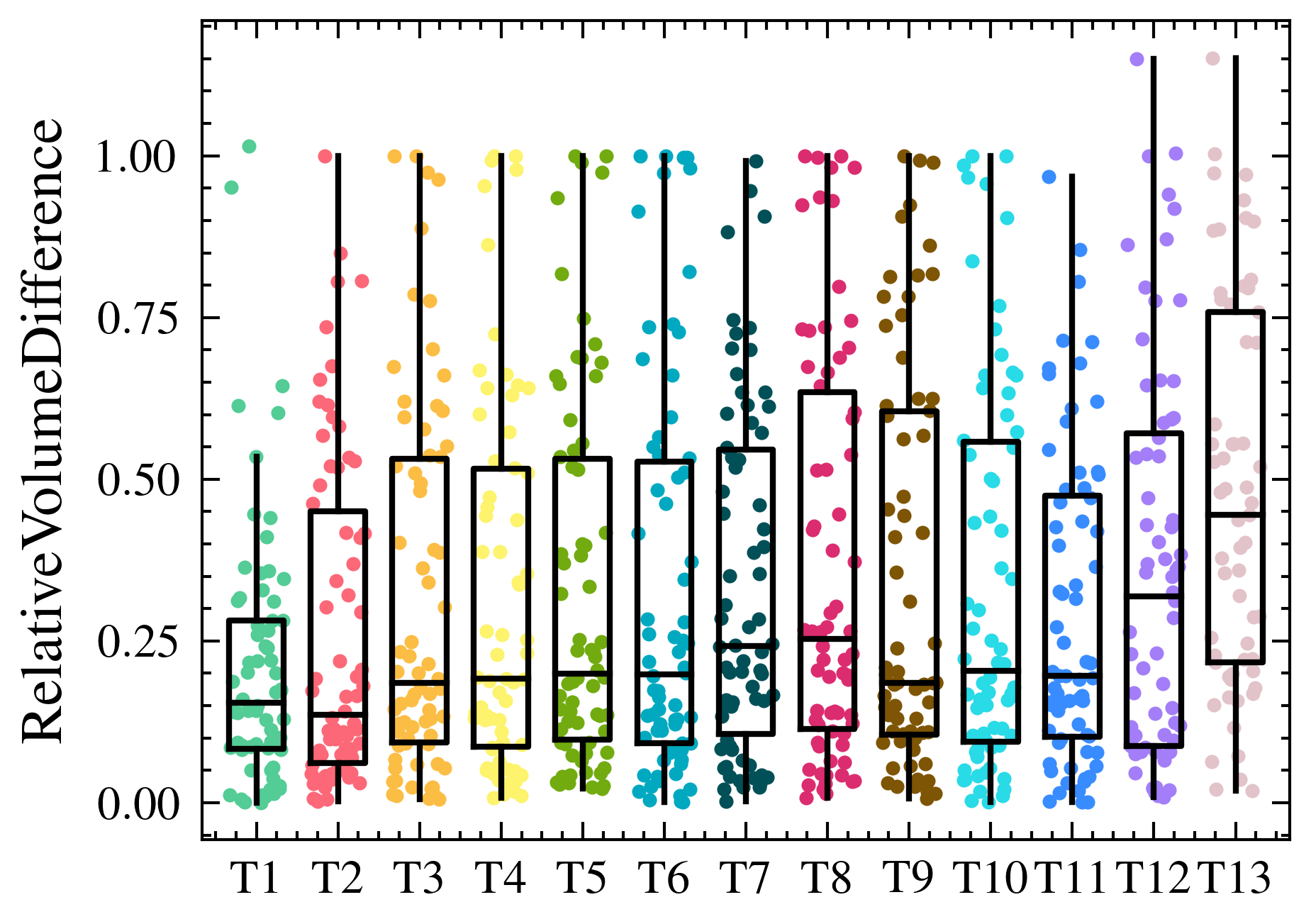}
		\label{subfig:RelativeVolumeDifference}
	}
	\subfigure[Hausdorff Distance]{
		\includegraphics[width=8cm]{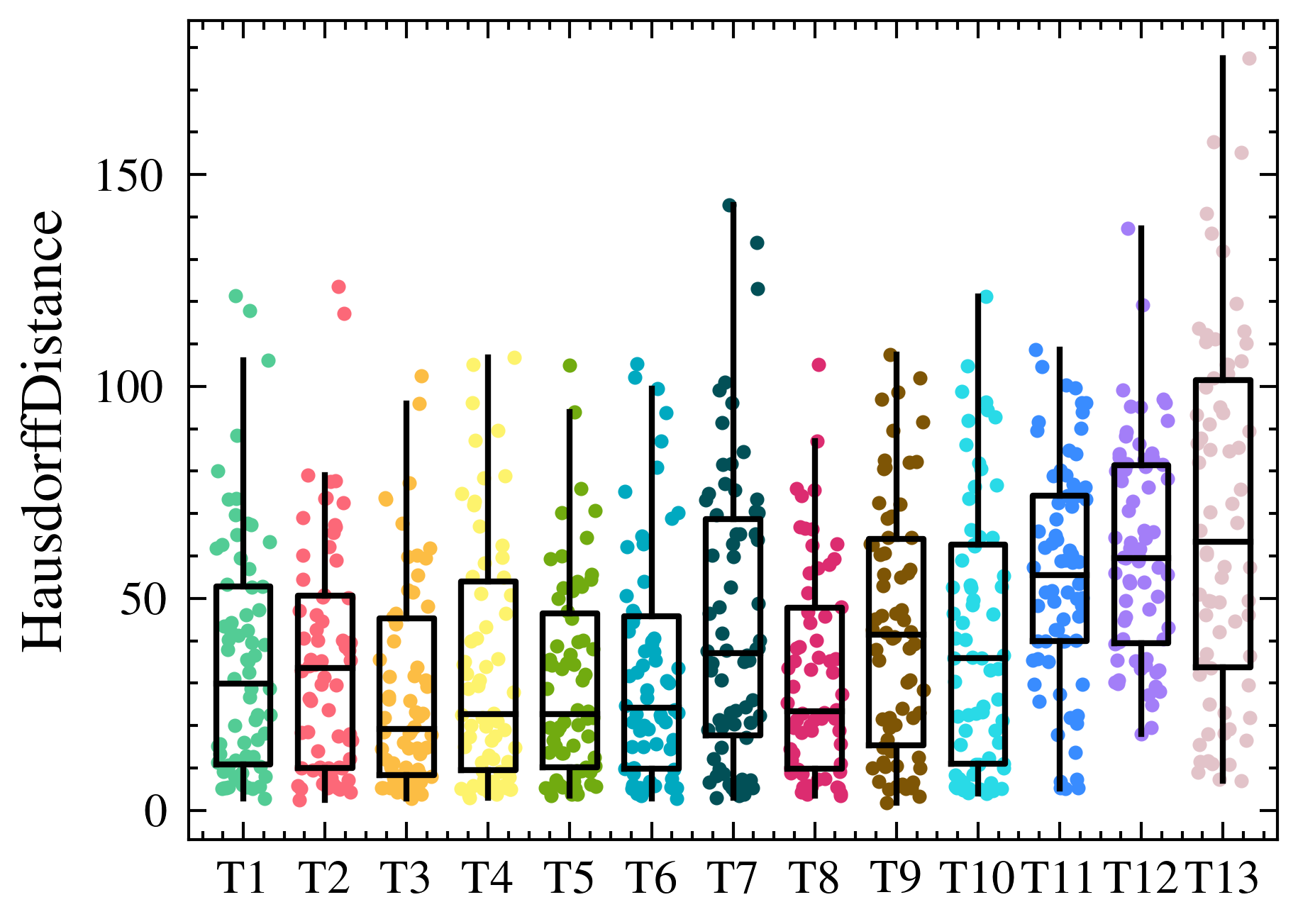}
		\label{subfig:HausdorffDistance}
	}
	\caption{Box plots of the experimental results on different evaluation metrics for all the submitted teams. The dots denote the individual scores of the 70 cases in the testing set.}
	\label{fig:results}
\end{figure*}
\subsection{Evaluation results and Analysis}
\subsubsection{Segmentation performance}\label{subsec:seg_results}
The segmentation performance of the baseline model and other participants' algorithms for validation and testing set are illustrated in Table. \ref{tab:challenge_results_validation} and Table. \ref{tab:challenge_results} respectively. In Table. \ref{tab:challenge_results}, we reported the average DSC, RVD, NSD and HD in the table, respectively. Our baseline model, SLEX-Net \cite{li2021hematoma} obtained a DSC score of 52.83$\%$. Most of other teams improved the baseline model in all four metrics. The average DSC score, RVD, NSD for the participants lies in [40.22$\%$,72.06$\%$], [0.21, 1.55], and [25.11$\%$, 53.59$\%$], respectively. The best results on DSC, RVD, and NSD metrics achieved only 72.06$\%$, 0.21, 53.59$\%$, respectively.  The overall performances are much lower than many other segmentation tasks, proving the great challenge of intracranial hemorrhage segmentation task. More importantly, most of the teams obtained 'infinite' for the averaged HD because their method mistakenly diagnosed some difficult ICH cases with tiny hemorrhages as normal subjects. The infinite results made it challenging to effectively rank the HD metric for different methods. In our challenge, we treat all ‘inf’ teams the same rank on HD which are inferior to others. Because we believe effectively diagnosis hard samples is also important in this task. 
Moreover, Fig. \ref{fig:results}(a)-(d) demonstrate the results distribution across all the subjects in the testing dataset with box plots. It can be inferred that the standard deviations of the results distribution for top ranking teams are smaller that that of lower ranking ones, and also fewer outliers exists for them as well.
\begin{figure}
	\centering
	\includegraphics[height=8.58cm,width=7cm]{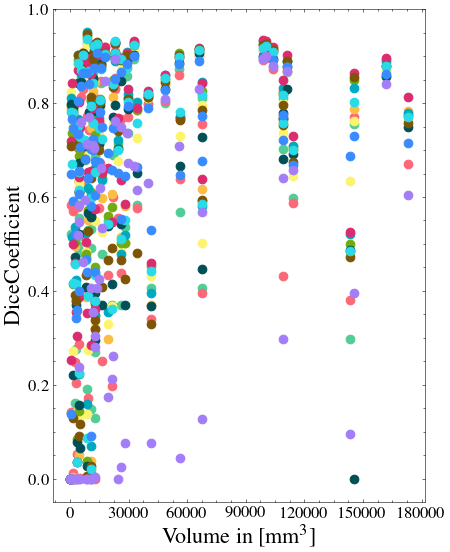}
	\caption[]{The relationship between different Dice coefficients and the hematoma volume sizes demonstrates that the cases with smaller hematoma volumes are hard cases.} 
	\label{fig:volumn_}
\end{figure} 
\begin{figure}
	\centering
	\includegraphics[height=6.477cm,width=9cm]{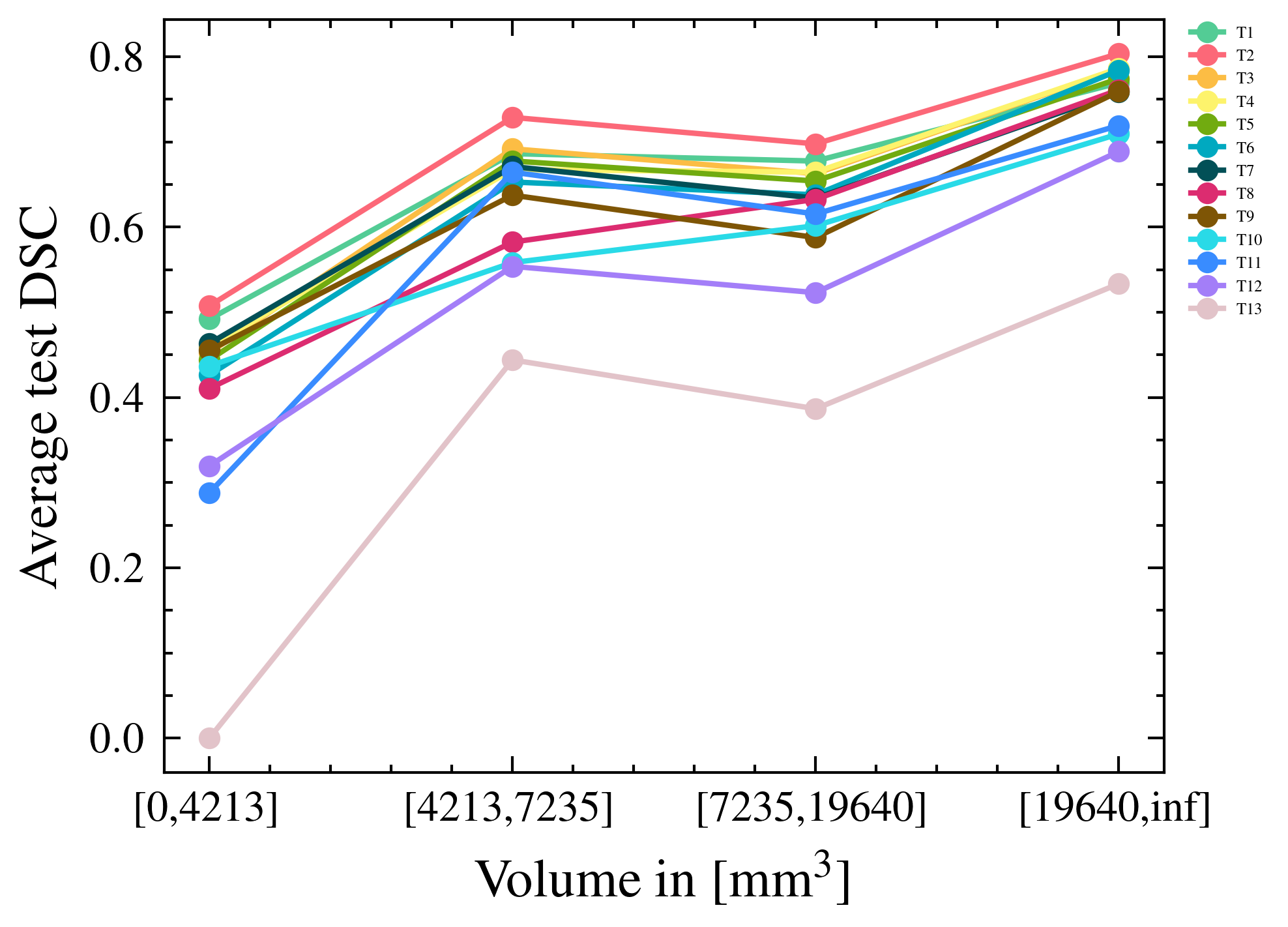}
	\caption[]{The team-wise 'Volumn-DSC' relationship figure shows that the DSC scores improve with the increase of volume sizes for different algorithms from the participants. It is generated by separating the 70 testing cases with four different volume size groups: including $[0,4213], [4213,7235],[7235,19640],[19640,inf]$, respectively, and the average DSC score was calculated based on the results in each group.} 
	\label{fig:volumn_dice}
\end{figure} 
\subsubsection{Hematoma Volume Analysis}\label{subsubsec:volumn_size}
In this section, we analyzed the relationship between hematoma volume size and the segmentation performances for different algorithms. The volume sizes of ICH are calculated by multiplying the voxel numbers of ICH and the pixel spacing in x,y,z dimensions, which is consistent with the method in \cite{ironside2020fully, li2021hematoma}.  Fig. \ref{fig:volumn_} highlights the correlation between volume size and the DSC scores with a scatter plot. It demonstrates that hemorrhages with small volume sizes are difficult to segment, while large hematoma ICHs are relatively easier to achieve better segmentation results. Fig. \ref{fig:volumn_dice} shows the segmentation performance for all the methods with four hematoma volume size groups. It is generated by separating the 70 testing cases with four different volume size groups: including $[0,4213], [4213,7235],[7235,19640],[19640,inf]$, respectively, and the average DSC score was calculated based on the results in each group. Fig. \ref{fig:volumn_dice} further proves that the DSC scores improve with the increase of volume sizes for different algorithms from the participants.
\begin{figure}
	\centering
	\includegraphics[height=8cm,width=8cm]{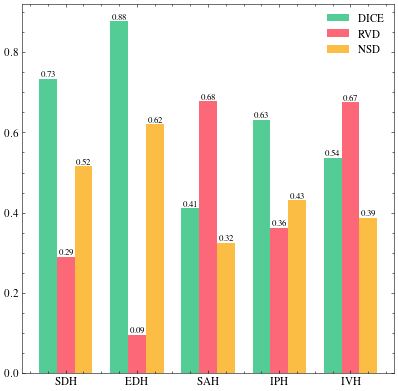}
	\caption[]{The bar chart on Dice Coefficient for different kinds of intracranial hemorrhages shows that SAH is the most difficult class to segment.} 
	\label{fig:dice_class_wise}
\end{figure} 
\subsubsection{Hemorrhage Sub-type Analysis}\label{subsec:sub_type_analysis}
Different sub-types of the intracranial hemorrhages are located at distinct positions of the brain, and patients can suffer from combinations of several kinds of hemorrhages. Certain types of hemorrhages usually present various different characteristics, leading to varied difficulties for distinguishing from normal brain tissues. Fig. \ref{fig:dice_class_wise} illustrates the average DSC value for different kinds of hemorrhages. It demonstrates that the SAH achieved the worst results in all metrics compared to other four kinds of ICHs. Hence, how to effectively segment SAH might be the most urgent problem needed to be solved to improve the ICH segmentation.   
\subsection{Challenge Ranking Analysis}
\begin{figure*}[htbp]
	\centering
	\subfigure[Significance map for Dice Coefficient]{
		\includegraphics[width=7cm]{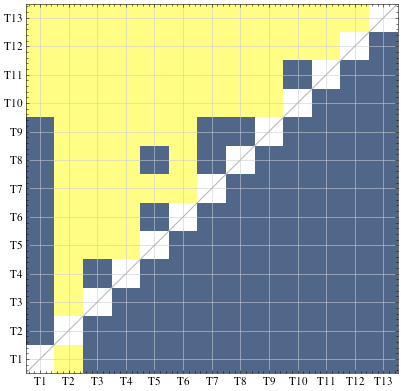}
		\label{subfig:significance_dice}
	}
	\subfigure[Significance map for Normalized Surface Dice]{
		\includegraphics[width=7cm]{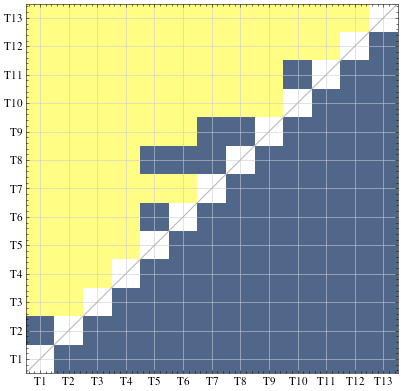}
		\label{subfig:significance_SurfaceDice}
	}
	\quad  
	\subfigure[Significance map for Relative Volume Difference]{
		\includegraphics[width=7cm]{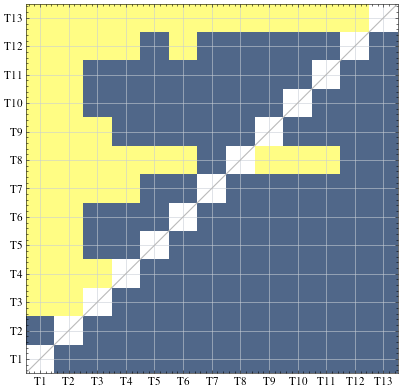}
		\label{subfig:significance_RelativeVolumeDifference}
	}
	\subfigure[Significance map for Hausdorff Distance]{
		\includegraphics[width=7cm]{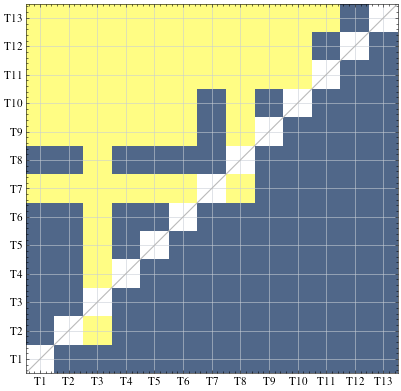}
		\label{subfig:significance_HausdorffDistance}
	}
	\caption{The significant superiority maps for ranking robustness analysis of different evaluation metrics. In each of the four maps, yellow blocks means that the evaluation metric for teams on the x-axis are significantly superior to those from the teams on the y-axis, which blue blocks means no significant superiority. The pairwise significant test with one-sided Wilcoxon signed rank test at 5\% significance level is adopted in our experiment.}
	\label{fig:significance_testing}
\end{figure*}

Similar to the significance analysis in many biomedical image segmentation challenges \cite{ma2022fast, oreiller2022head}, we utilized the significance map to demonstrate the pairwise significant superiority between different algorithms,  as is illustrated in Fig. \ref{fig:significance_testing}. Specifically, we choose to perform significant test with one-sided Wilcoxon signed rank test at $5\%$ significance level. In Fig. \ref{fig:significance_testing} (a-d), most of the yellow blocks are above the diagonal and the blue blocks are under the diagonal, indicating that most of the teams with smaller rank are significantly superior to those with larger ranks. Moreover, it also shows that different metrics have distinct ability to distinguish the good and bad performances among different algorithms. For example, the DSC, NSD and HD of T7 are significantly superior to that of T12, however, there exists no significant superiority on RVD metric.  
\section{Discussions}	
\subsection{2D/3D architecture Choice}
The algorithm summary in section \ref{subsec:algorithm_summary} shows that the participants chose different algorithm implementations for 2D or 3D methods. We noticed that the winner method adopted the 2D/3D combination method, and most of the 3D methods outperformed the 2D implementations, yet we cannot draw definite conclusions on which kinds of methods are superior to another since there are many other factors contributing to the final results. However, we believe that directly utilizing 2D networks would lose significant context information among slices, which has been proved in numerous medical image segmentation tasks \cite{chen2016combining,zheng20183,dou2016multilevel,li2021hematoma}. Therefore, how to effective incorporate inter-slice contextual information would be a fundamental problem for improving ICH segmentation. To this end, many participants utilized 3D UNet implementation, however, this might not be the optimal solution considering that the CT volumes in this challenge are anisotropic (pixel spacing: $0.42mm\times0.42mm\times5mm$) \cite{liu20183d}, thus more effective techniques for exploiting inter-slice context for anisotropic volumes are needed. 
\subsection{Bottlenecks for ICH segmentation}
The hematoma volume analysis in section \ref{subsubsec:volumn_size} demonstrates the inferior segmentation performance for hemorrhages with small volume sizes. The degradation of the segmentation indicates that the hemorrhage cases with small volume sizes are hard to segment. Fig. \ref{fig:volumn_} shows that all the methods proposed by the participants have trouble dealing with very small hemorrhages. The majority of the cases that achieve a DSC score lower than 0.3 are those subjects with hemorrhage volume smaller than 15000$m^3$, and the overall DSC performances for all the subjects significantly deteriorate with substantial low DSC scores. Therefore, one important bottleneck for ICH segmentation is the small hemorrhage lesion segmentation, and effectively resolving this problem would certainty improve the overall segmentation performance and achieve better ranking in the challenge. Besides, the hemorrhage sub-type analysis in section \ref{subsec:sub_type_analysis} shows that the subarachnoid hemorrhage (SAH) achieved the worst results in all metrics, with average DSC score for only 0.41. Thus, another bottleneck for ICH segmentation is how to deal with the subarachnoid hemorrhage. In conclusion, the future directions for the researches of ICH segmentation may be concentrated on the above-mentioned two bottlenecks. The researches of the hemorrhage diagnosis would be greatly improved by resolving these extremely hard cases.
\subsection{Evaluation Metrics Analysis}
We highly suggest the use of DSC, NSD and the RVD as the evaluation metrics for the ICH segmentation benchmark. According to the descriptions in section \ref{subsec:eval_metrics}, and section \ref{subsec:seg_results}. The HD and NSD are similar metrics that are used to evaluate the discrepancy between the target and predicted boundaries. However, we came across multiple extreme cases with average HD metrics equal to infinite when the predicted methods mistakenly diagnosed those hard cases with small hemorrhage lesions as normal head scans. The 
infinite values make it challenging to effectively rank different algorithms on that metric. However, the NSD metric has the same upper bound as DSC (100\%), and there will be no such circumstances occur. Therefore, Hausdorff distance might not be a good metric choice for the INSTANCE challenge, and we consider abandoning it in the future INSTANCE challenges. 
\subsection{Limitations and Future work}
Although this year's INSTANCE challenge has achieved great success with numerous participants around the world, it still suffers from lots of limitations. They are mainly consist of three aspects:
\subsubsection{Data collection and annotation}
Even though the INSTANCE2022 challenge has provided a relatively large dataset, they are mainly collected from a single institution with the same CT scanner. Although it could work in our challenge, it would definitely restrict the generalization of the model developed by different participants. In addition, for the data annotation, we only delineate the hemorrhage regions as foreground without considering the ICH sub-types, which are actually important information in clinical diagnosis and can also guide the segmentation of ICH. 
\subsubsection{Task designs}
In this years' INSTANCE challenge, we only consider the hemorrhage segmentation task. However, it is also significant to perform ICH classification and hematoma volume quantification, which are highly clinical-related. The design of multiple tasks would simultaneously make the challenge more comprehensive and provide more diverse research directions for the participants. In conclude, we will enhance the single-task challenge to a multi-task one in the future challenges. 
\subsubsection{Source code Availability}
In this years' INSTANCE challenge, we highly recommended the participants to make their implementations to the public, and didn't make it a mandatory option. As a result, we only find out one team make their code available. We didn't demand them to share the code because we don't expect it to be an obstacle for participating in this challenge. However, we notice that the code is too significant to be ignored for promoting the development in this research field. Therefore, we consider making it mandatory for top participants to make their code public available for future INSTANCE challenges. 

\subsubsection{Future works for INSTANCE}
We are currently working to promote the INSTANCE 2022 Challenge in many different aspects. Detailed improving directions are as follows:
\begin{itemize}
	\item \textbf{More multi-institutional data.} We will collect more ICH data from different CT scanner and different hospitals to improve the generalization of methods that are trained based on the INSTANCE benchmark. 
	\item \textbf{More annotations and comprehensive task designs.} We will annotate the different ICH sub-types of each CT scans and also calculate the hematoma volume of each cases to provide more clinical-related datasets. Meanwhile, based on the above-mentioned extra annotations, we further expand the single-task challenge to a multi-task one, which simultaneously performs hemorrhage segmentation, classification and volume quantification tasks.  
	\item \textbf{Mandatory options for open-source code.} To promote the advancement of the intracranial hemorrhage diagnosis, the top participants in the future INSTANCE challenge are required to share their code to the public. 
\end{itemize}

\begin{figure*}
	\centering
	\includegraphics[height=16cm,width=16cm]{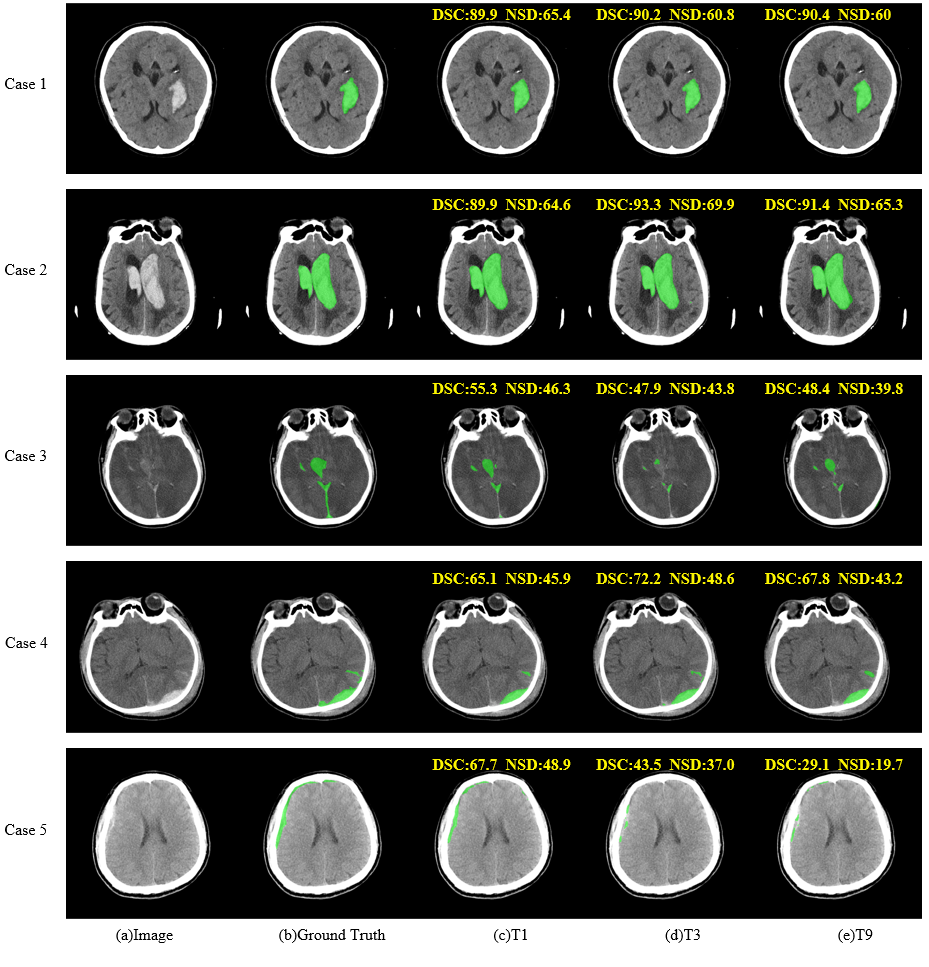}
	\caption[]{Segmentation results for different ICH sub-types in terms of DSC and NSD scores. The blue color denotes the ICH lesion.} 
	\label{fig:qualitative_results}
\end{figure*} 

\section{Conclusion}
The INSTANCE challenge provides a novel benchmark for objectively measuring different intracranial hemorrhage segmentation methods in non-contrast head CT scans. A total of 13 teams successfully submitted their methods, and the winner solution achieved a DSC score of 0.6925 on the testing set, dramatically improving our baseline network. We have made the training set, the methodology descriptions and evaluation code public available on the challenge website, we hope this would promote the development in the ICH segmentation field. 
The challenge is now remains open for post-challenge submissions via Grand Challenge platform for benchmarking further algorithm exploitation. In the future, we will collect more multi-institutional data to improve the generalization of methods that are trained on the benchmark, and also perform more clinical-relevant annotations on ICH sub-type and hematoma volumes and expand the single-task challenge to a multi-task one.


\section*{Acknowledgments}
We sincerely appreciate all the members in INSTANCE2022 organization team for their hard work. Without your continuous devotion to this challenge, it would not be that successful. This work was supported by the National Natural Science Foundation of China under Grant 62001144, 62272135 and 62001141, and by Science and Technology Innovation Committee of Shenzhen Municipality under Grant RCBS20210609103820029 and JCYJ20210324131800002.

\appendix[]\label{sec:appendix}

\quad In (Li and Chen, 2022), Li and Chen used a combination of nnU-Net and uncertainty estimation ensemble strategy. Their experiments showed that even though the 2D nnU-Net could not achieve the overall dice accuracy of 3D nnU-Net, it performed better results than 3D nnU-Net when the intracranial hemorrhage had very small area or blurred boundaries. Therefore, they use both 2D and 3D nnU-Net to predict the final result.  Furthermore, in order to further alleviate the segmentation issue of small area intracranial hemorrhage and maintain stability during training, they utilized the weighted cross-entropy loss to replace simple cross-entropy loss in the nnU-Net. Due to the unbalanced intracranial hemorrhage types and intracranial hemorrhage areas, the models trained in different folds might predict completely different results. Simply average the predicted results from the models provide no additional benefit for these cases. To this end, they propose a simple but efficient uncertainty estimation ensemble strategy.  For those cases with high uncertainty values, they use the voting method to get the final result. Use nnU-Net's own data augmentation methods.

\quad In (Siddiquee et al., 2022 \cite{siddiquee2022automated}), Siddiquee et al. used the 2D version of encoder-decoder backbone based on with an asymmetrically larger encoder to extract image features and a smaller decoder to reconstruct the segmentation mask\footnote{Implementation: https://monai.io/apps/auto3dseg}. For the encoder part, they used 5 stages of down-sampling and 2D ResNet blocks that each block’s output is followed by an additive identity skip connection. Furthermore, they used batch normalization and ReLU. For the decoder part, the decoder structure is similar to the encoder one, but with a single block per each spatial level. Each decoder level begins with upsizing with transposed convolution. In the preprocessing, they applied random rotation and random zoom on each axis with a probability of 0.4 and random contrast adjustment and random Gaussian noise with a probability of 0.2. The random coarse shuffle and random flips were applied on each axis with a probability of 0.5. In the training, they randomly split the entire dataset into 5-folds and trained a model for each. Moreover, they used L2 norm regularization on the convolution kernel parameters with a weight of $1e^{-5}$. The DiceCE loss is used for training.

\quad In (Sanner and Mukhopadhyay, 2022), Sanner and Mukhopadhyay used nnU-Net for the segmentation and propose an evaluation of contour-based losses. Specifically, they integrated both the Hausdorff-distance loss as proposed by \cite{karimi2019reducing} and the contour loss proposed by \cite{jurdi2021surprisingly}. While the former estimates the Hausdorff distance, the latter extracts the contour of both the prediction and the ground truth and minimizes the mean square error between them. In practice, Dice loss and CE loss were used as loss function and the Hausdorff-distance loss or the contour loss was used depending on the experiment. Furthermore, rather than using the standard z-normalization of nnU-Net for input images, they chose to clip the intensity values to [0 - 100]. A five-fold cross-validation was used to train five models and all models were ensembled to make the final prediction. The "insane DA" scheme was used for data augmentation.

\quad In (Zhao et al., 2022), Zhao et al. used two stage  3D cascade U-Net network for ICH segmentation. For the stage 1, the basic module of the encoder and decoder is Conv-Instance Norm-LeakyReLU\cite{zhang2017dilated}. The operation of downsampling in the encoder is achieved by max pooling. The upsampling operation in the decoder is achieved by using the transpose convolution of 2 ×  2 ×  2. For the stage 2, a 3D U-Net was cascaded to the model, whose input is the output of probability map of the first stage.  The 5-fold cross-validation was used for the training. In the preprocessing, the HU of CT images were clipped according to three different windows and levels, and corresponding range of HU were [0, 80], [-20, 180] and [-150, 230]. The intensity of the voxel above the range were assigned the value of upper limit in range, and the intensity below the range is assigned the value of lower limit in range. Then the three images with different HU range clip were served as three channels and treated as one image.

\quad In (Zhang and Ma, 2022), Zhang and Ma used the standard nnU-Net.First, a 3D U-Net processes downsampled data, the resulting segmentation maps are upsampled to the original resolution.Then, these segmentations are concatenated as one-hot encodings to the full resolution data and refined by a second 3D U-Net. The preprocessing includes cropping,resampling and normalization. Meanwhile, random rotation, random scaling, random elastic transformation, gamma correction, and mirror were used to augment the data. The 3D nnU-Net was trained with an weighted combination of Dice loss and cross-entropy loss.  The results on the test set were obtained as an ensemble of five models.

\quad In (Liu et al., 2022 \cite{liu2022voxels}), Liu et al. used an ensemble model that combined viola-Unet and nnU-Net networks\footnote{Implementation: https://github.com/samleoqh/Viola-Unet}. For the viola-Unet, it relies on voxels in feature space that intersect along orthogonal levels to provide an attention U-Net, which is an asymmetric encoder-decoder architecture with 7-depth layers.  Overall, the Viola module is composed of three key blocks, i.e., the adaptive average pooling (AdaAvgPool) module that squeezes the input feature volume into three latent representation spaces along each axis of the input feature patch. The customized dense dilated convolutions merging (DDCM) networks fuses cross-channel and non-local contextual information on each orthogonal direction with adaptive kernel sizes, dilated ratios and strides. The Viola unit constructs the voxels intersecting along orthogonal level attention volume based on fused and reshaped cross-channel-direction latent representation spaces. They trained all networks with randomly sampled patches of fixed size as input and applied a combination loss function of the dice loss and Focal loss for all their experiments. In preprocessing, CT image and ground truth labels were reoriented into ”RAS” format, then resized to a standard spacing of 1×1×5 mm3 using trilinear interpolation for the image and nearest-neighbor interpolation for the label. Each CT image was windowed into three image intensity ranges, and re-scaled to the range [0, 1] by min-max normalization and then stacked as 3-channel volumes to serve as inputs with the (C, H, W, D) shape, and then the 3-channel 3D volume was normalized on only non-zero values with calculated mean and std on each channel separately. The data augmentations include random crop, random zoom, Gaussian noise, Gaussian smooth, rotation, random shift, random scale, flips, random contrast. Furthermore, they manually select the best prediction on each validation example from each submission as the pseudo-label and put them into our training set to fine-tune our models repeatedly in practice.

\quad In (Liang, 2022), Liang proposed a nnUNet-based method for 3-dimensional intracranial hemorrhage segmentation. In the preprocessing, the authors first deal with the data in method windowing and decide to choose a width of 59 and a center of 96 for the image windowing by experiment. After windowing, in order to arrange the information of image, the author used a threshold to ensure the gray value of the image in a certain standard interval, unified data input. Then, downsampling the X and y axes, normalize the spacing of the slice axis to Slice\_down\_scale. A sampling includes maximum sampling, average sampling, summation area sampling, and random area sampling. Finally, nnU-Net does the rest of the preprocessing. In the training, the author uses CE loss + DICE loss as loss function.  Furthermore, to deal with category imbalance, oversampling was used, with 66.7\% of the samples coming from random locations in the selected training sample's, while 33.3\% of the patches were guaranteed to contain one of the foreground classes present in the selected training sample (randomly selected). The number of foreground patches was rounded to force a minimum value of 1 (resulting in one random patch and one foreground patch with a batch size of 2). Use nnU-Net's own data augmentation methods.

\quad In (Geiger et al., 2022), Geiger et al. used classic U-Net architecture and the network was conducted with the jax version of the e3nn library which enables the creation of neural networks equivariant to translations, rotations, and mirroring. Specially, the convolution kernels in the original architecture are replaced by a 3D e3nn voxel convolution of diameter 5 mm. Furthermore, they used three 2x2x2 downsampling operations which halve the resolution in the encoding path and three corresponding trilinear upsampling operations on the decoding path. A Gaussian error linear unit activation function and instance normalization was used after each convolution. For the preprocessing, each CT volume was windowed to three different Hounsfield unit value ranges, scaled, and added to a separate channel which served as the model input. To increase the variety in the data, a random diffeomorphic deformation was performed on each training sample. The loss function employed was cross-entropy loss. Eight models were trained, each on 80 randomly sampled subsets from the training dataset. The final prediction was performed by applying each of the eight models to patches of size 144x144x13 with padding discarding 22x22x2 pixels on each side, a sliding window with an overlap of 26 pixels and Gaussian weighing, and then averaging the model outputs. For the final prediction, they take the ensemble average of the eight models.

\quad In (Qayyum et al., 2022), Qayyum et al. developed a coarse and fine segmentation model for intracranial hemorrhage segmentations. They trained two different models for intracranial hemorrhage segmentations. In the first model, they trained 2DDensNet for coarse segmentation and cascaded the coarse segmentation masks output in the fine segmentation model along with input training samples. The proposed model is implemented made by a dense encoder followed by a non-dense decoder. The dense encoder consists of 5 dense blocks, each consisting of 6 dense layers followed by a transition layer. Each dense layer consists of 2 convolutional layers with batch normalization and ReLU activation functions.  The model is trained using 5-fold cross-validation. To compute the final prediction, 2D images are stacked to make a 3D segmentation mask. The predicted segmentation mask is further cascaded in a fine segmentation model. In the fine stage, they used the nnU-Net model with fivefold cross-validation. The binary cross-entropy function was used as loss function. HorizontalFlip (p=0.5), VerticalFlip (p=0.5), and RandomGamma (p=0.8) were used to augment the dataset for training the proposed model. In addition, the dataset is normalized between 0 and 1 using the max and min intensity normalization method. The training shape of each volume is fixed (256x256x16) and resample the prediction mask to the original shape for each validation volume using the linear interpolation method.

\quad In (Abramova et al., 2022), Abramova et al. used an approach based on a 3D U-Net architecture which incorporates squeeze-and-excitation blocks that similarly to their previous work \cite{abramova2021hemorrhagic}. For the preprocessing, coil removal and skull stripping were used, and a symmetric image was created for each case by flipping the original non-contrast CT and registering it to the initial one using the FLIRT algorithm from the FSL toolbox. For the normalization of input images, they performed percentile based range adjustment and used 0.5 and 99.5 percentiles of brain-related voxels for clipping together with image-based calculated mean and standard deviation normalization. For the issue of class imbalance, they used a balanced sampling patch extraction technique, where we extracted an equal number of patches representing both classes from each image. Specifically, to avoid extracting a lot of patches from image background, they restricted the area to extract the negative patches within the brain mask and set a target number of patches to extract from each image in the training set. Half of them are uniformly extracted from the brain tissue area and represent negative class, while the other half is extracted from the lesion voxels. They augment the proposed dataset by choosing difficult cases and adding them into the training set again, meanwhile performing flipping and rotation, ensuring that more difficult patches are generated for training. The Dice loss and cross-entropy loss was used as loss function. To prevent overfitting, they used early stopping technique when approaching the minimal loss on validation set. The five-fold cross-validation strategy was used for the training. For the validation and testing stages, an ensemble with all the 5 models obtained in the cross-validation experiment was used to generate the final prediction masks. The probability masks obtained from the 5 models were averaged and thresholded to obtain the final binary mask for each case. Considering the results on the validation set, postprocessing was added to their pipeline to reduce the number of false positives. Specifically, as sizes of lesions vastly vary in the provided images, they remove all the lesions with the volume less than 10\% of the biggest one in the post-processed image.

\quad In (Montagnon and Letourneau-Guillon, 2022), Montagnon and Letourneau-Guillon used an ensemble approach including the Attention U-Net and SegResNet (with or without variational autoencoder) architectures combined with different loss functions. Specifically, they trained U-Net and SegResNet separately to use different loss functions including combinations of Dice with either Cross-Entropy loss or Focal loss, Tversky loss and Generalised Dice loss. Then leveraging all predictions, an ensemble voting approach allowed prediction of a final volume. Finally, to further remove potential false positive predictions, predicted clusters were filtered by preserving ones with a volume larger than 36 pixels, an elevation above or equal to 3 slices and a mean density within [40; 80] HU range. In the preprocessing, in order to assess hemorrhage properties, they used DBSCAN, a density-based clustering algorithm, in order to extract connected pixels corresponding to hemorrhagic areas in each exam. Then they clipped images in the range [-10; 140] HU. Taking into account the intracranial hemorrhage subtype distribution in the training dataset, they using Euler transforms consisting of rotations of either -$\frac{\pi}{2}$ or -$\frac{\pi}{2}$ around z-axis and translations ranging from -30 to 30 pixels, 10 pixels stepwise for subarachnoid and subdural hemorrhage subtypes images. Considering the limited size of the dataset, they used random orthogonal rotations and cropping for images in the training phase. In order to limit class imbalance issues, models were trained only on images containing at least one pixel of positive class. All models were trained using original images size (i.e. 512 $\times$ 512), clipped within [-10;140] and divided by the range of considered densities, which is 150 in their configuration.

\quad In (Roca et al., 2022), Roca et al. used a simple 2D Unet-like model and trained it with a binary cross-entropy loss. Especially, the model input is a layer that performs the clipping operation between [0, 256] and a normalization between [-0.5, 3.5] directly inside the model. In the preprocessing, they clipping the HU intensities in the soft tissue range of interest. For the data augmentation, they performed rotations and mirroring in the axial plane, plus some amount of intensity shift. Due to the data stratification was based on the presence of a segmentation on a given slice (positive cohort) vs. absence of segmentation (negative cohort), they used during training a balanced 50\% / 50\% of each cohort per mini-batch. 

\quad In (Sindhura et al., 2022), Sindhura et al. proposed a deep learning framework which involves clinical knowledge and used U-Net3+ network for the segmentation. Specifically, they proposed a new data augmentation approach that leverages from the clinical knowledge that the two hemispheres of the human brain exhibit approximate symmetry. Due to the brain is approximately divided into two equal hemispheres by the midsagittal plane (MSP). So they use the MSP flipped versions of the CT scans as extra data. To extract MSP, they first apply the sobel edge detection method followed by thresholding to obtain the outline of the skull. An initial plane of reference is chosen to be the exact middle slice in the sagittal direction. A similarity metric is computed between the two hemispheres that are divided with the plane of reference. The reference plane is rotated by an angle of $\pm0.5^\circ$. The plane which yields maximum similarity is the required MSP. Furthermore, to improve the robustness of the model, the usual data augmentations such as shear, rotation, zoom, flip, elastic transform, noise etc are being used. In view of there exists a very high class imbalance between the hematoma and non-hematoma pixels. So only the slices which contain hemorrhages are used in the training process and all slices of each scan are tested in the testing phase. In addition, to differentiate between the hemorrhage region and skull bone, which share similar intensities, they have performed skull stripping on each scan for both the training and testing process. The sum of focal loss and Dice similarity loss is used as the loss function in the training process.

\bibliographystyle{IEEEtran}
\bibliography{reference}

\end{document}